\RequirePackage{fix-cm}
\documentclass{svjour3}
\smartqed  
\AtBeginDocument{%
  \setlength{\oddsidemargin}{\dimexpr(\paperwidth-\textwidth)/2-1in}%
  \setlength{\evensidemargin}{\oddsidemargin}%
  \setlength{\topmargin}{%
    \dimexpr(\paperheight-\textheight)/2-\headheight-\headsep-1in}%
}

\usepackage{latexsym}
\usepackage{amsmath}
\usepackage[colorlinks=true,linkcolor=blue,urlcolor=blue,citecolor=red]{hyperref}
\usepackage[utf8]{inputenc}
\usepackage{amsmath,amssymb,calc,amsfonts}

\font\tenscr=rsfs10 scaled1100
\font\sevenscr=rsfs7 
\font\fivescr=rsfs5 
\skewchar\tenscr='177
\skewchar\sevenscr='177
\skewchar\fivescr='177
\newfam\scrfam
\textfont\scrfam=\tenscr
\scriptfont\scrfam=\sevenscr
\scriptscriptfont\scrfam=\fivescr

\def\scri{{\fam\scrfam I}}
\def\scrm{{\fam\scrfam M}}

\begin{document}

\title{Sources of Gravitational Waves in Asymptotically flat Einstein-Maxwell Spacetime}


\author{G. D. Quiroga}


\institute{G. D. Quiroga \at
              Grupo de Investigaci\'on en Relatividad y Gravitaci\'on, Escuela de F\'isica, Universidad Industrial de Santander, A. A. 678, Bucaramanga, Colombia \\
              \email{gquiroga@uis.edu.co}           
}

\date{Received: date / Accepted: date}

\maketitle

\begin{abstract}
In this work, the dynamic of isolated systems in general relativity is described when gravitational radiation and electromagnetic fields are present. In this construction, the asymptotic fields received at null infinity together with the regularized null cone cuts equation, and the center of mass of an asymptotically flat Einstein-Maxwell spacetime are used. A set of equations are derived in the low speed regime, linking their time evolution to the emitted gravitational radiation and to the Maxwell fields received at infinity. These equations should be useful when describing the dynamic of compact sources, such as the final moments of binary coalescence and the evolution of the final black hole. Additionally, we compare our equations with those coming from a similar approach given by Newman, finding some differences in the motion of the center of mass and spin of the gravitational system.
\end{abstract}

\section{Introduction}\label{intro}
In September 2015 the first detection of gravitational waves from colliding Black Holes was made by LIGO \cite{LIGO2016}. This experimental confirmation of the gravitational radiation existence generated a strong impact on the scientific community. Close binary coalescence of black holes (BHs) and neutron stars (NSs) are the prime candidates for LIGO in the search for gravitational waves. These binary systems can be studied as systems where the energy, the linear and angular momentum are given off and carried away by the radiation. Although for most astrophysical processes one can give a Newtonian definition of these variables, in these very energetic situations a relativistic definition must be given. The problem lies in the impossibility of defining locally these variables and it is for this reason that only global definitions are due. Fortunately, using the notion of asymptotic flatness together with the inclusion of a 3-dim null boundary, called Null Infinity, one can give suitable definitions of the Bondi 4-momentum $P$ and (with considerable more debate among specialists) the mass dipole-angular momentum two form.

As mentioned, it is very important for these compact sources of gravitational radiation to define the notion of center of mass since the loss of momentum from the system causes a recoil or kick velocity to be imparted to the center of mass of the coalesced binary. In a recent work together with Kozameh \cite{KQ2}, a new definition of center of mass and spin for isolated sources of gravitational radiation was given, its time evolution was derived and a comparison with analogous formulae coming from the post-Newtonian approximation was given. In that work, a slow motion approximation was taken since most of the time the astrophysical sources do not acquire relativistic velocities as a result of the gravitational radiation emission. However, the contributions of the electromagnetic fields in the equation of motion of the center of mass and spin were neglected.

The electromagnetic fields play an important role in gravitational theories even in General Relativity (GR). There are many astrophysical sources which are modeled as charged \cite{adamo2014kerr} or magnetized \cite{harding2006physics,romani1990unified} compact sources, even there are models of galaxies where the electrodynamic contributions are so significant in its behavior and stability \cite{gutierrez2013conformastatic,cardona2016analytical}. In this context, the coalescence and fusion of binary systems is not an exception, especially in systems such NS-NS, where the magnetic moments present in neutron stars influences upon the dynamics of, and resulting gravitational waves \cite{anderson2008magnetized}. The electromagnetic contributions are significant even in other astrophysical sources, for example, in the gravitational wave emission by a distorted rotating fluid star \cite{bonazzola1996gravitational}, where the distortion of stochastic magnetic fields might lead to gravitational waves detectable by the VIRGO or LIGO interferometers. Then, it is clear that electromagnetic fields and the gravitational waves should be included in the description of any astrophysical systems. In the particular case of the center of mass and spin, the electromagnetic field causes a change in its dynamics and evolution, thus it is very important to include these corrections in our previous work to get a more accurate description of the final state and merge of the binary coalescence.

In this paper, we extend our previous results, including the electrodynamics contributions in the equation of motion of the center of mass, spin, and the Bondi 4-momentum. These equations are derived assuming the slow motion approximation for the center of mass velocity. This is not a limitation on the applicability of our formalism since in most astrophysical processes only a fraction of the total mass is lost as gravitational radiation. In other words, even if two coalescing stars (or black holes) are approaching each other at relativistic speeds, if the center of mass is initially at rest it will not acquire a relativistic velocity. So, in these scenarios, where the gamma factor for the center of mass velocity is close to one, these equations can be applied.

Many works were written on this topic, among the most recent we can mention those published by Newman and collaborators \cite{kozameh2009electrodynamic,kozameh2005electromagnetic,ANK} where definitions of quantities such as the complex center of charge, and complex center of mass was given. However, our work updates and correct the equations obtained previously by these authors using a new definition for the center of mass and using a real equation (versus a complex) to describe the cuts at null infinity. These ingredients incorporate some extra terms in the final equations adding the electrodynamic corrections in the equation of motion, and interaction terms between electromagnetic dipoles and gravitational radiation.

The article is organized as follows, in Section 2 we give the needed mathematical constructions. In particular, we introduce the notion of asymptotically flat spacetimes and derive a set of null vectors in the neighborhood of null infinity. Also, we introduce the Newman-Penrose formalisms and all the quantities needed in the following sections. In Section 3 we present a brief review of our previous work presenting the special two-form that defines the dipole mass moment and total angular momentum for an isolated source.  We also introduce the so called regularized null cone cuts as the special family of cuts that are used to define the center of mass and the center of charge. Additionally, in this section we derive the main results obtaining the relationships between these global variables together with their time evolution. In section 4 we compare our results with those obtained by Newman, and in section 5 we give some application introducing a simple model of a Gravitational Charged Particle, and we show that using our equations the classical gyromagnetic ratio for a charged black hole is obtained. Finally, we conclude this work with some remarks and conclusions.

\section{Asymptotically flat spacetime}\label{sec:1}
The notion of asymptotically flat spacetime is an adequate tool to analyze the gravitational and electromagnetic radiation coming from an arbitrary compact source.
A spacetime  $(\scrm, g_{ab})$  is called asymptotically flat if the curvature tensor vanishes as it approaches infinity along the future directed null geodesics of the spacetime. The geometrical notion of an asymptotically flat spacetime can be formalized by the following definition \cite{ntod}.

\emph{Definition:} a future null asymptote is a manifold $\hat \scrm$ with boundary $\scri^+ \equiv \partial \hat\scrm$ together with a smooth Lorentzian metric $\hat{g}_{ab}$, and a smooth function $\Omega$ on $\hat\scrm$ satisfying the following
\begin{itemize}
\renewcommand\labelitemi{$\circ$}
  \item $\hat{\scrm}=\scrm \cup \scri^+$
  \item On $\scrm$, $\hat{g}_{ab}=\Omega ^2 g_{ab}$ with $\Omega >0$
  \item At $\scri^+$, $\Omega=0$, $n_a \equiv \partial _a \Omega \neq 0$ and $\hat {g}^{ab}n_a n_b =0$
\end{itemize}
We assume $\scri^{+}$ to have topology $S^2\times R$. The two-surface metric becomes
\begin{equation}\label{metricasphy}
ds^2= -\frac{4r^2 d\zeta d\bar\zeta}{P^2},
\end{equation}
Now, with the usual choice of $\Omega=r^{-1}$ as the conformal factor, (\ref{metricasphy}) gives the induced metric of $\scri^+$,
\begin{equation} \label{metrics2}
d\hat{s}^2=-\frac{4d\zeta d\bar \zeta}{P^2}.
\end{equation}
here $P(u,\zeta,\bar \zeta)$ is a strictly positive arbitrary function.

In the neighborhood of $\scri^+$ it is possible to introduce a Newman-Unti  system (NU) \cite{nu}, with coordinates  $(u, r, \zeta ,\bar\zeta )$. In this system, the time $u$ represents a family of null surfaces, $r$ is the affine parameter along the null geodesics of the constant $u$ surfaces, and $(\zeta,\bar\zeta)$ are the complex stereographic angle that labels the null geodesics of the null surface. Associated with the NU coordinates, there is a null tetrad system denoted by $(l_{a}^{\ast}$,$n_{a}^{\ast}$,$m_{a}^{\ast}$,$\bar{m}_{a}^{\ast})$. These vectors are defined to satisfy the following conditions \cite{ntod}
\begin{align}
l_{a}^{\ast}&=\nabla _{a}u,\\
l^{a \ast} n_a^{\ast}&=-m^{a\ast}\bar{m}_a^{\ast}=1,\\
l^{a \ast}m_a^{\ast}=l^{a \ast}\bar{m}_a^{\ast}&=n^{a \ast}m_a^{\ast}=n^{a \ast}\bar{m}_a^{\ast}=0,\\
l^{a \ast}l_a^{\ast}=n^{a \ast}n_a^{\ast}&=m^{a \ast}m_a^{\ast}=\bar{m}^{a \ast}\bar{m}_a^{\ast}=0.
\end{align}
The indices can be raised and lowered using the global metric $g_{ab}$ which can be written as
\begin{equation} \label{metrica}
g_{ab}=l^{\ast}_{a}n^{\ast}_{b}+n^{\ast}_{a}l^{\ast}_{b}-m^{\ast}_{a}\bar{m}^{\ast}_{b}-\bar{m}^{\ast}_{a}m^{\ast}_{b}
\end{equation}
Now, we will introduce the Bondi coordinates which are a particular choice of a NU system. In a Bondi system, the coordinates $(u_B, r_B, \zeta, \bar\zeta)$ are related with the NU coordinates as follows
\begin{align}
u_B&=Z(u,\zeta,\bar\zeta) \\
r_B&=Z^{\prime}r
\end{align}
The first equation come from one of the tetrad freedom \cite{ntod}, which allow to introduce a different choice of the original cut $u=const$, this freedom implies that
\begin{equation}
u=T(u_B,\zeta,\bar\zeta).
\end{equation}
$Z$ and $T$ are smooth real function, where $T$ is the inverse of $Z$. These functions satisfy that $Z^{\prime}\dot{T}=1$, here the dot and the prime means derivative with respect $u_B$ and $u$ respectively.

Also, a quantity $\eta$ that transforms as $\eta \rightarrow e^{is\lambda }\eta$ under a rotation $m^{a\ast} \rightarrow e^{i\lambda }m^{a\ast }$ is said to have a spin weight $s$. For any function $f(u, \zeta, \bar{\zeta})$, we define the differential operators $\eth^{\ast }$ and $\bar{\eth }^{\ast }$ \cite{ANK} by
\begin{eqnarray}
\eth^{\ast } f&=&P^{1-s}\frac{\partial (P^{s}f)}{\partial \zeta }, \label{eth}\\
\bar{\eth }^{\ast }f&=&P^{1+s}\frac{\partial (P^{-s}f)}{\partial \bar{\zeta}}, \label{ethb}
\end{eqnarray}
where $f$ has a spin weight $s$ and $P$ is the conformal factor of the metric (\ref{metrics2}). The particular choice $P_0=(1+\zeta \bar{\zeta})$ makes the two-surface metric (\ref{metricasphy}) a sphere. Furthermore, using $P=P_0 Z^{\prime}$ in eqs. (\ref{eth}-\ref{eth}) and remaining $u=const$, one can write the transformation from a topological sphere to unit sphere as
\begin{eqnarray}
\eth^{\ast }f&=&Z^{\prime }\eth f+sf\eth Z^{\prime } \\
\bar{\eth }^{\ast }f&=&Z^{\prime }\bar{\eth }f-sf\bar{\eth }Z^{\prime },
\end{eqnarray}
where the symbol ${}^{\ast}$ distinguishes between the differential operators applied with conformal factor $P$ and $P_0$ respectively at $u=const$.
The above equation will be used below to expand regular functions on the sphere in the standard spherical harmonic basis.

In the same way, for the Bondi system, a set of four null vectors can be introduced $(l^a,n^a,m^a,\bar{m}^a)$ and expand one of those bases in terms of the other. In this case, we write the NU as a linear combination of the Bondi null vectors just making the choice $l_{a}^{\ast }=\nabla _{a}T(u_B,\zeta ,\bar{\zeta})$ and using the orthogonality of the null vectors as follows
\begin{eqnarray}
l_{a}^{\ast } &=&\frac{1}{Z^{\prime }}[l_{a}-\frac{L}{r_{B}}\bar{m}_{a}-%
\frac{\bar{L}}{r_{B}}m_{a}+\frac{L\bar{L}}{r_{B}^{2}}n_{a}], \label{la}\\
n_{a}^{\ast } &=&Z^{\prime }n_{a}, \label{na}\\
m_{a}^{\ast } &=&m_{a}-\frac{L}{r_{B}}n_{a} \label{ma}, \\
\bar{m}_{a}^{\ast } &=&\bar{m}_{a}-\frac{\bar{L}}{r_{B}}n_{a}, \label{mba}
\end{eqnarray}
where
$$L(u_{B},\zeta ,\bar{\zeta })=\eth Z(u,\zeta ,\bar{\zeta }). $$

\subsection{The Newman-Penrose formalism}\label{sec:12}
In the Newman-Penrose formalism, one introduces twelve complex spin coefficients, five complex functions encoding Weyl tensor, three complex Maxwell scalars, and ten functions encoding Ricci tensors in the tetrad basis \cite{np}. These complex functions are the primary quantities used in the asymptotic formulation of the General Relativity.

In this article, we will focus on the general form of the asymptotically flat solutions of the Einstein-Maxwell equations, starting with the Ricci rotation coefficients in term of the Bondi coordinates introduced in the previous section, this means that all introduced functions depend on these coordinates. Now, the Ricci rotation coefficients $\gamma_{\mu \nu \rho}$ can be written as
\begin{equation}
\gamma_{\mu \nu \rho}={\lambda ^{a}}_{\rho}{\lambda ^{b}}_{\nu}\nabla _{a}\lambda_{b \mu},
\end{equation}
where $\mu,\nu,\rho=1,2,3,4$ are tetrad indexes and
\begin{equation}
{\lambda^a}_{\mu}=(l^a,n^a,m^a,\bar{m}^a).
\end{equation}
Also, the Ricci rotations coefficients satisfy
\begin{equation}
\gamma _{\mu \nu \rho}=-\gamma _{\nu \mu \rho}.
\end{equation}%
Then, the spin coefficients are defined as combinations of the $\gamma_{\mu \nu \rho}$ such as
\begin{eqnarray}
\alpha  &=&\frac{1}{2}(\gamma _{124}-\gamma _{344}),\quad \lambda =-\gamma
_{244},\quad \kappa =\gamma _{131},  \nonumber  \label{spincoef} \\
\beta  &=&\frac{1}{2}(\gamma _{123}-\gamma _{343}),\quad \mu =-\gamma
_{243},\quad \rho =\gamma _{134}, \\
\gamma  &=&\frac{1}{2}(\gamma _{122}-\gamma _{342}),\quad \nu =-\gamma
_{242},\quad \sigma =\gamma _{133},  \nonumber \\
\varepsilon  &=&\frac{1}{2}(\gamma _{121}-\gamma _{341}),\quad \pi =-\gamma
_{241},\quad \tau =\gamma _{132}.  \nonumber
\end{eqnarray}
Since the spacetime is assumed to be empty in a neighborhood of the null infinity, the gravitational field is given by the Weyl tensor. Using the available tetrad one defines five complex scalars
\begin{eqnarray}
\psi _{0}=-C_{abcd}l^{a}m^{b}l^{c}m^{d}; \quad \psi _{1}=-C_{abcd}l^{a}n^{b}l^{c}m^{d}, \nonumber\\
\psi _{2}=-\frac{1}{2}(C_{abcd}l^{a}n^{b}l^{c}n^{d}-C_{abcd}l^{a}n^{b}m^{c}{\bar{m}}^{d}), \\
\psi _{3}=C_{abcd}l^{a}n^{b}n^{c}{\bar{m}}^{d}; \quad \psi _{4}=-C_{abcd}n^{a}{\bar{m}}^{b}n^{c}{\bar{m}}^{d}, \nonumber
\end{eqnarray}
When an electromagnetic field is present, we can introduce the Maxwell tensor $F_{ab}=\partial_a A_b-\partial_b A_a$,  from where we compute three complex Maxwell scalars given by
\begin{equation}
\phi _{0}=F_{ab}l^{a}m^{b}; \quad \phi _{1}=\frac{1}{2}F_{ab}(l^{a}n^{b}+m^{a}{\bar{m}}^{b}); \quad \phi _{2}=F_{ab}n^{a}{\bar{m}}^{b},
\end{equation}
Finally the Peeling theorem of Sachs \cite{sachs} tell us about the asymptotic behaviors of the Weyl and the Maxwell scalars, which are the following
\begin{align}
\psi _{0}&=\psi _{0}^{0}r_B^{-5}+O(r_B^{-6}), \quad \psi _{1}=\psi _{1}^{0}r_B^{-4}+O(r_B^{-5}), \quad \psi _{2}=\psi _{2}^{0}r_B^{-3}+O(r_B^{-4}) \nonumber\\
\psi _{3}&=\psi _{3}^{0}r_B^{-2}+O(r_B^{-3}), \qquad \psi _{4}=\psi _{4}^{0}r_B^{-1}+O(r_B^{-2}), \\
\phi _{0}&=\phi _{0}^{0}r_B^{-3}+O(r_B^{-4}), \quad \phi _{1}=\phi _{1}^{0}r_B^{-2}+O(r_B^{-3}), \quad \phi _{2}=\phi _{2}^{0}r_B^{-1}+O(r_B^{-2}). \nonumber
\end{align}
the Peeling also tell us about the behavior of the spin coefficients \cite{ANK},
\begin{align}
\kappa  &=\pi =\varepsilon =0;\qquad \rho =\bar{\rho};\qquad \tau =\bar{%
\alpha}+\beta   \nonumber  \label{spincoef-des-r} \\
\rho  &=-r_B^{-1}-\sigma ^{0}\bar{\sigma}^{0}r_B^{-3}+O(r_B^{-5})  \nonumber \\
\sigma  &=\sigma ^{0}r_B^{-2}+[(\sigma ^{0})^{2}\bar{\sigma}^{0}-\psi
_{0}^{0}/2]r_B^{-4}+O(r_B^{-5})  \nonumber \\
\alpha  &=\alpha ^{0}r_B^{-1}+O(r_B^{-2})  \nonumber \\
\beta  &=\beta ^{0}r_B^{-1}+O(r_B^{-2}) \\
\gamma  &=\gamma ^{0}-\psi _{2}^{0}(2r_B^{2})^{-1}+O(r_B^{-3})  \nonumber \\
\mu  &=\mu ^{0}r_B^{-1}+O(r_B^{-2})  \nonumber \\
\lambda  &=\lambda ^{0}r_B^{-1}+O(r_B^{-2})  \nonumber \\
\nu  &=\nu ^{0}+O(r_B^{-1})  \nonumber
\end{align}
where the relationships among the r-independent functions are given by
\begin{eqnarray}
\alpha ^{0} &=&-\bar{\beta}^{0}=-\frac{\zeta }{2},\qquad \gamma ^{0}=\nu
^{0}=0,   \nonumber \\
\omega ^{0} &=&-\bar{\eth }\sigma ^{0}, \qquad \lambda ^{0}=\dot{\bar{\sigma}}^{0},\qquad \mu ^{0}=-1,  \nonumber
\end{eqnarray}
with $\sigma^0$ the value of the Bondi shear at null infinity, this complex scalar is called the Bondi free data since $\ddot{\sigma}^0$ yields the gravitational radiation reaching null infinity. Although the NP formalism is the basic working tool for our analysis, we simply have given an outline of this formulation and leave the reference \cite{ntod,np} for extra details. 
Additionally, we can define the Weyl scalars in NU using the fact that the Weyl tensor ${C_{abc}}^d$ is conformally invariant \cite{ntod}.
\begin{align}
\psi _{1}^{\ast} &={C_{abc}}^d n^{a\ast}l^{b\ast}l^{c\ast}m_{d}^{\ast}\simeq \psi _{1}^{0\ast}r^{-4},\\
\sigma^{\ast} &=m^{\ast a}m^{\ast b}\nabla _{a}l_{b}^{\ast }\simeq\sigma ^{0\ast}r^{-2}.
\end{align}
From equations (\ref{la}-\ref{mba}) we can find transformations from NU to Bondi for any scalar or spin coefficient \cite{AN,KQ}. In particular we are interested in
\begin{equation}\label{transformacion}
\frac{{\psi }_{1}^{0\ast}}{Z^{\prime 3}}=[\psi _{1}^{0}-3L\psi _{2}^{0}+3L^{2}\psi_{3}^{0}-L^{3}\psi _{4}^{0}],
\end{equation}
where ${\psi }_{1}^{0\ast}$ is constructed from the NU tetrad. Similarly, we find the relation between $\sigma^{0\ast}$ and $\sigma^{0}$ \cite{AN}
\begin{equation}\label{sigma}
\frac{\sigma^{0\ast }}{Z^{\prime}}=\sigma ^{0}-\eth^{2}Z,
\end{equation}
where $\sigma^{0\ast}$ is the NU shear \cite{nu}. In the same way for the electromagnetic field we have,
\begin{equation}\label{maxwell}
\frac{\phi _{0}^{0\ast }}{Z^{\prime 2}}(u,\zeta ,\bar{\zeta })=[\phi _{0}^{0}-2L\phi_{1}^{0}+L^{2}\phi _{2}^{0}](u_B,\zeta ,\bar{\zeta }).
\end{equation}
Finally, we introduce the evolution equations, which are given by the following Bianchi identities \cite{ntod}
\begin{eqnarray}
\psi_{2}^{0}-\bar{\psi }_{2}^{0} &=&\bar{\eth }^{2}\sigma ^{0}-\eth ^{2}\bar{\sigma }^{0}+\bar{\sigma }^{0}\dot{\sigma }^{0}-\sigma ^{0}\dot{\bar{\sigma }}^{0}, \label{psi2}\\
\psi_{3}^{0}&=&\eth \dot{\bar{\sigma }}^{0},\\
\psi_{4}^{0}&=&-\ddot{\bar{\sigma }}^{0}.
\end{eqnarray}
These equations are valid up to second order in $\sigma^{0}$. Form Eq. (\ref{psi2}) we can define the so called mass aspect \cite{ANK}
\begin{equation} \label{aspmasa}
\Psi =\psi _{2}^{0}+\eth ^{2}\bar{\sigma }^{0}+\sigma ^{0}\dot{\bar{\sigma }}^{0},
\end{equation}
which satisfies the following reality condition
\begin{equation}
\Psi =\bar{\Psi },
\end{equation}
using the asymptotic Bianchi identities \cite{ntod}, we obtain the most important differential equations for this work
\begin{align}
\dot{\psi}_{1}^{0}&=-\eth \Psi +\eth ^{3}\bar{\sigma }^{0}+\eth \sigma ^{0}\dot{\bar{\sigma }}^{0}+3\sigma ^{0}\eth \dot{\bar{\sigma} }^{0}+\frac{4G}{c^4}\phi _{1}^{0}\bar{\phi }_{2}^{0}, \label{psi1prima}\\
\dot{\psi_{2}^{0}}&=-\eth ^{2}\dot{\bar{\sigma }}^{0}-\sigma ^{0}\ddot{\bar{\sigma }}^{0}+\frac{2G}{c^4}\phi _{2}^{0}\bar{\phi }_{2}^{0}, \label{bianchi2} \\
\dot{\phi}_0^0&=-\eth \phi_1^0 +\sigma^0 \phi_2^0,\label{bianchi4} \\
\dot{\phi}_1^0&=-\eth \phi_2^0. \label{bianchi3}
\end{align}
Combining the mass aspect $\Psi$ with $\dot{\psi}_{2}^{0}$, the Bianchi identities $\dot{\psi}_{2}^{0}$ can be rewritten in the form
\begin{equation}
\dot{\Psi}=\dot{\sigma} ^{0}\dot{\bar{\sigma }}^{0}+\frac{2G}{c^4}\phi _{2}^{0}\bar{\phi }_{2}^{0}. \label{psiprima}
\end{equation}
Note that the r.h.s of the Bianchi identities are only functions of $\sigma^0$, $\Psi$ and the Maxwell scalars $\phi_1^0$ and $\phi_2^0$.
In addition, we can introduce the Bondi mass and three-momentum through the equations \cite{ntod}
\begin{align}
M &=-\frac{c^{2}}{2\sqrt{2}G}\int \Psi d\Omega  \\
P^{i}&=-\frac{c^{3}}{6G}\int {\Psi }l^{i}d\Omega.
\end{align}

\section{Spin and Center of Mass Review}
The concept of spin and the center of mass in general relativity is of great importance, although the general construction of the Kozameh-Quiroga approach can be found in refs. \cite{KQ,KQ2}. In this section, we will outline only the most relevant aspects of these previous works.

\subsection{Regularized null cone cuts}\label{sec:RNC}
In flat spacetime, the null cone cuts or NC cuts for short, are smooth surfaces that can be written as regular functions on the sphere,
\begin{equation}\label{flat_cut}
Z_0 = x^a \ell_a, \qquad x^a=(R^0,R^i) \qquad \ell_a=(Y^0_0,-\frac{1}{2}Y_{1i}^0)
\end{equation}
with $x^a$ the apex of the null cone and where $Y^0_0,Y_{1i}^0$ the $\ell=0,1$ tensorial spin-s harmonic \cite{ngilb}. If the apex $x^a(u)$ describes a timelike worldline in Minkowski space, the NC cuts describe a one parameter foliation of null infinity.

Now, to generalize this concept for asymptotically flat spacetimes, the first difficulty to be solved is that a generic NC cut is not a smooth two-surface at null infinity since the NC cut have self-intersections and caustics. However, using the regularized null cone cuts (RNC) equation for the smooth function $Z$, it is always possible to find a neighborhood at null infinity where a NC cut is a smooth two-surface (see Appendix A of ref. \cite{KQ2}). The RNC cuts correspond to the linear regular solutions of the NC equation \cite{BKR}. One expects that the leading contribution to the solution comes from the Huygens part of the RNC cuts equation, which can be written as,
\begin{equation}\label{RNC_cuts}
{\bar \eth}^2 \eth^{2}Z={\bar \eth}^2 \sigma^0(Z,\zeta,{\bar \zeta}) + \eth^2{\bar \sigma}^0(Z,\zeta,{\bar \zeta}).
\end{equation}
This linearized version was independently derived by L. Mason \cite{Mason}, and by Fritelli et al. \cite{FKN2}. Since (\ref{RNC_cuts}) only contains $\ell \geq 2$, the kernel of (\ref{RNC_cuts}) is a four-dimensional space $x^a$, i.e. a flat cut $Z_0 = x^a \ell_a$.

The RNC cut equation yield monoparametric families of NU cuts whose areas are time dependent and in general are not unit spheres. It should be noted that the center of mass is accelerated by the emission of gravitational radiation and, in consequence, the null cuts performed at null infinity by the center of mass worldline will correspond to a family of NU cuts. Now, the solution to the RNC cuts equation can be found using the perturbative solution
\begin{equation}
Z=Z_0+Z_1+Z_2+...,
\end{equation}
where each term in the series is determined from the previous one and the free data $\sigma^0(u_B,\zeta,{\bar \zeta})$. The first two terms satisfy
\begin{eqnarray}
{\bar \eth}^2 \eth^{2}Z_0&=&0,\\
{\bar \eth}^2 \eth^{2}Z_1&=&{\bar \eth}^2 \sigma^0(Z_0,\zeta,{\bar \zeta}) + \eth^2{\bar \sigma}^0(Z_0,\zeta,{\bar \zeta}),\label{linear cut}
\end{eqnarray}
Clearly, the zeroth order term $Z_0$ is simply the flat cut (\ref{flat_cut}) and it has been assumed that in the absence of radiation the Bondi shear vanishes. The first non-trivial term of eq. (\ref{RNC_cuts}) is found by solving (\ref{linear cut}), which is given by
\begin{eqnarray}
Z_{1}=R^{0}-\frac{1}{2}R^{i}Y_{1i}^{0}+\left( \frac{\sigma _{R}^{ij}}{12}+\frac{%
\sqrt{2}}{72}{\sigma}_{I}^{\prime ig}R^{f}\epsilon ^{gfi}\right) Y_{2ij}^{0} \label{Z1}
\end{eqnarray}
Note that $Z_1$ depends on the real and imaginary parts of the Bondi shear, also if $x^a(u)$ describes any worldline, then $Z_i$ describes a NU foliation up to the order needed. This is what one would expect in a perturbation expansion since the imaginary part of the Bondi shear is related to the current quadrupole moment, but the real part comes from the mass quadrupole moment \cite{Kidder}.
Now, equation (\ref{RNC_cuts}) can be compared with the good cut equation
\begin{equation}\label{good_cut}
\eth^2Z_C=\sigma^0(Z_c,\zeta,\bar\zeta),
\end{equation}
The good cut equation yields complex cuts with vanishing shear, while the RNC cuts equation yields NU cuts whose shear depends linearly on the Bondi shear. Thus, from the point of view of available structures at null infinity, we could start with the RNC cuts equation (\ref{RNC_cuts}). On its four-dimensional solution space, one constructs a Lorentzian metric following the NSF procedure \cite{FKN2}. A perturbative solution gives a Minkowski space together with flat cuts (\ref{flat_cut}) at its lowest order, and the linearized RNC cuts (\ref{Z1}) at first order.

\subsection{Spin and Center of Mass Definitions}
The center of mass-angular momentum tensor is defined from the Winicour-Tamburino linkage integral \cite{Wini,LMN}. The mass dipole moment and the angular momentum can be defined using the real and the imaginary part of the linkage integral as follows \cite{KQ2}),
\begin{eqnarray}\label{DJ}
D^{\ast i}+ic^{-1}J^{\ast i}=-\frac{c^{2}}{12\sqrt{2}G}\left[ \frac{2\psi _{1}^{0}-2\sigma
^{0}\eth \bar{\sigma}^{0}-\eth(\sigma ^{0}\bar{\sigma}^{0})}{Z^{\prime 3}}\right]^{\ast i}. \nonumber\\
\end{eqnarray}
We assume that a special worldline $x^a(u)$ exist, where the mass dipole moment $D^{\ast i}|_{u=const}$ vanishes for each $u=const.$ cut. This special worldline on this special RNC foliation, will be called the ``Center of Mass worldline'' of the system. The angular momentum $J^{i\ast}$ evaluated at the center of mass will be the intrinsic angular momentum $S^i$. Thus, the center of mass worldline $x^a(u)$ will correspond with the worldline following by the center of mass in the solution space of the RNC equation (\ref{RNC_cuts}), this worldline is then determined using the equation,
\begin{equation}
Re\left[ \frac{2\psi _{1}^{0}-2\sigma^{0}\eth \bar{\sigma}^{0}-\eth(\sigma ^{0}\bar{\sigma}^{0})}{Z^{\prime 3}}\right]^{\ast i}=0. \label{CoM}
\end{equation}
Since the 4-velocity of the worldline is normalized to one (using the spacetime metric), we will use this norm to fix the timelike component of the worldline coordinate. Thus, the only freedoms left are the spatial components of the worldline $R^i$. The above equation gives three algebraic equations from which these components are obtained. Additionally, the intrinsic angular momentum is given by
\begin{equation}
S^{i}=-\frac{c^{3}}{12\sqrt{2}G}Im\left[ \frac{2\psi _{1}^{0}-2\sigma
^{0}\eth \bar{\sigma}^{0}-\eth(\sigma ^{0}\bar{\sigma}^{0})}{Z^{\prime 3}}\right]^{\ast i}. \label{Si}
\end{equation}
Now, to describe the motion of the center of mass will be use a simpler system, such a Bondi system, where there is a precise notion of mass and momentum.
For that, it is important to introduce some analogous quantities to $D^{\ast i}$ and $J^{\ast i}$ in a Bondi tetrad system, these quantities are defined as
\begin{equation}\label{DJB}
D^{i}+ic^{-1}J^{i}=-\frac{c^{2}}{12\sqrt{2}G}\left[ 2\psi _{1}^{0}-2\sigma
^{0}\eth \bar{\sigma}^{0}-\eth(\sigma ^{0}\bar{\sigma}^{0})\right]^{i}.
\end{equation}
The natural way to link these two definitions is finding the transformation between the quantities $(\psi _{1}^{0\ast }, \sigma^{0\ast}, \eth)$ in NU to $(\psi _{1}^{0}, \sigma^{0}, \eth_{B})$ in a Bondi system, for that we use eqs. (\ref{transformacion}) and (\ref{sigma}) to write
\begin{eqnarray}
D^{\ast i}(u) &=&D^{i}(u_B)+\frac{3c^{2}}{6\sqrt{2}G}Re[\eth Z(\Psi -\eth ^{2}\bar{\sigma}^{0})+F]^{i}\label{Dexp} \\
J^{i\ast }(u) &=&J^{i}(u_B) +\frac{3c^{3}}{6\sqrt{2}G}Im[\eth Z(\Psi -\eth ^{2}\bar{\sigma}^{0})+F]^{i}\label{Jexp}
\end{eqnarray}
with
\begin{eqnarray}\label{F}
F&=&-\frac{1}{2}(\sigma ^{0}\eth \bar{\eth }^{2}Z+\eth ^{2}Z\eth \bar{\sigma}^{0}-\eth ^{2}Z\eth \bar{\eth }^{2}Z)\nonumber\\
&&-\frac{1}{6}(\bar{\sigma}^{0}\eth ^{3}Z+\bar{\eth }^{2}Z\eth \sigma ^{0}-\bar{\eth }^{2}Z\eth ^{3}Z).
\end{eqnarray}
Before continuing we will introduce some assumptions and approximations that facilitate the calculations of the final equations.
First, we assume that $\sigma=0$ at some initial time, this assumption fixes the supertranslation freedom.
We also consider the quadratic terms in the gravitational radiation and we assume that the Bondi shear only has a quadrupole term.
Finally, we consider $R^i$ as a small deviation from the coordinates origin and $R^0 = u$ assuming the slow motion approximation.
Following these works assumptions, we can write the $Z_1$ solution (\ref{Z1}) of the RNC cut as,
\begin{align}
Z_1&= u+ \delta u \nonumber\\
\delta u &\equiv -\frac{1}{2}R^{i}(u)Y_{1i}^{0}(\zeta,\bar \zeta)+\frac{1}{12}\sigma _{R}^{ij}(u)Y_{2ij}^{0}(\zeta,\bar \zeta ), \label{Z1apr}
\end{align}
and making a Taylor expansion up to first order in $\delta u$ and its derivatives we have
\begin{eqnarray}
D^{\ast i}(u) &=&D^{i}(u)+[D^{\prime}(u)\delta u]^{i}+\frac{3c^{2}}{6\sqrt{2}G}Re[\eth Z(\Psi -\eth ^{2}\bar{\sigma}^{0})+F]^{i}\nonumber \\
&=&D^{i}(u)+\frac{3c^{2}}{6\sqrt{2}G}Re[\eth Z(\Psi -\eth ^{2}\bar{\sigma}^{0})+F]^{i}+\frac{c^{2}}{6\sqrt{2}G}Re[(\eth \Psi -\eth ^{3}\bar{\sigma}^{0})\delta u]^{i}, \nonumber\\ \label{Dest_exp}
\end{eqnarray}
and
\begin{eqnarray}
J^{\ast i}(u) &=&J^{i}(u)+[J^{\prime}(u)\delta u]^{i}+\frac{3c^{3}}{6\sqrt{2}G}Im[\eth Z(\Psi -\eth ^{2}\bar{\sigma}^{0})+F]^{i},\\
&=&J^{i}(u)+\frac{c^{3}}{6\sqrt{2}G}Im[(\eth \Psi -\eth ^{3}\bar{\sigma}^{0})\delta u]^{i}+\frac{3c^{3}}{6\sqrt{2}G}Im[\eth Z(\Psi -\eth ^{2}\bar{\sigma}^{0})+F]^{i},\nonumber\\
\label{Jest_exp}
\end{eqnarray}
where in the r.h.s. of eqs. (\ref{Dest_exp}) and (\ref{Jest_exp}), and based on the above assumptions we have used that $D^{i \prime}(u)+icJ^{i \prime}(u) \approx [\dot {D}^{i}(u_B)+ic\dot{J}^{i}(u_B)]|_{u_B= u}$. Now, inserting the following tensorial spin-s harmonics expansion \cite{ngilb},
\begin{eqnarray}
\sigma^0 &=&\sigma ^{ij}(u)Y_{2ij}^{2}(\zeta,\bar \zeta ),\nonumber \\
\psi _{1}^{0} &=&\psi _{1}^{0i}(u)Y_{1i}^{1}(\zeta,\bar \zeta )+\psi_{1}^{0ij}(u)Y_{2ij}^{1}(\zeta,\bar \zeta ),\nonumber \\
\Psi  &=&-\frac{2\sqrt{2}G}{c^{2}}M-\frac{6G}{c^{3}}P^{i}(u)Y_{1i}^{0}(\zeta,\bar \zeta )+\Psi ^{ij}(u)Y_{2ij}^{0}(\zeta,\bar \zeta ),\nonumber\\
\phi _{0}^{0} &=&\phi _{0}^{0i}(u)Y_{1i}^{1}(\zeta ), \label{exp} \\
\phi _{1}^{0} &=&Q+\phi _{1}^{0i}(u)Y_{1i}^{0}(\zeta ),\nonumber \\
\phi _{2}^{0} &=&\phi _{2}^{0i}(u)Y_{1i}^{-1}(\zeta ),\nonumber
\end{eqnarray}
where $M$ is the Bondi mass and $P^i$ is the Bondi momentum \cite{ntod}.  Finally, in the center of mass worldline the mass dipole moment vanishes, i.e $D^{\ast}=0$, and $J^{\ast i}=S^i$, so eqs. (\ref{Dest_exp}) and (\ref{Jest_exp}) will be reduced to
\begin{align}
MR^{i}&=D^{i}+\frac{8}{5\sqrt{2}c}\sigma^{ij}_RP^j, \label{momdip}\\
J^{i}&=S^{i}+R^{j}P^{k}\epsilon ^{ijk}.\label{angmomentum}
\end{align}
Since the center of mass and the angular momentum was defined using only the Weyl scalar $\psi_1^0$ and $\sigma^0$, these definitions and the subsequent equations are independent of the electromagnetic fields.

\subsection{Electromagnetic Contributions to the Center of Mass}\label{sec:2.3}
The time evolution of $D^i$ and $J^i$ follows from the Bianchi identity \textcolor[rgb]{1.00,0.00,0.00}{for} $\psi_1^0$, where we must insert the proper factor of $\sqrt{2}$ to account for the retarded time, $u_{ret} = \sqrt{2} u_B$. The use of the retarded time, $u_{ret}$, is important in order to obtain the correct numerical factors in the expressions for the final physical results \cite{ANK}. Note that the two last eqs. (\ref{momdip}) and (\ref{angmomentum}) remain unchanged in term of $u_{ret}$ or $u_B$. From now on, we will adopt  the symbol ``prime'' for $\partial_{u_{ret}}$. Now, starting from the definition (\ref{DJB}) and replacing the real and imaginary $l=1$ component of (\ref{psi1prima}) we can  write,
\begin{eqnarray}
D^{i \prime}&=&P^{i}-\frac{1}{3\sqrt{2}c}(\phi_{1R}^{0k}\phi _{2I}^{0j}+\phi _{1I}^{0j}\phi _{2R}^{0k})\epsilon ^{ijk}-\frac{Q}{3c}\phi _{2R}^{0i},\label{Ddot}\\
J^{i \prime }&=&\frac{c^{3}}{5G}(\sigma _{R}^{kl}{\sigma}_{R}^{jl\prime}+\sigma
_{I}^{kl}{\sigma}_{I}^{jl\prime})\epsilon ^{ijk}-\frac{1}{3\sqrt{2}}(\phi
_{1R}^{0k}\phi _{2R}^{0j}+\phi _{1I}^{0j}\phi _{2I}^{0k})\epsilon ^{ijk}+%
\frac{Q}{3}\phi _{2I}^{0i}.\nonumber\\ \label{Jdot}
\end{eqnarray}
Note that the second term in (\ref{Jdot}) corresponds to the electromagnetic contribution to the angular momentum (see ref \cite{KQ}).
In the same way, we can use the Bianchi identity (\ref{psiprima}) to obtain the flux laws for the Bondi mass and linear momentum,
\begin{eqnarray}
M^{\prime}&=&-\frac{c}{10G}({\sigma}_{R}^{ij\prime}{\sigma}_{R}^{ij\prime}+%
{\sigma}_{I}^{ij\prime}{\sigma}_{I}^{ij\prime})-\frac{1}{6c}(\phi
_{2R}^{0i}\phi _{2R}^{0i}+\phi _{2I}^{0i}\phi _{2I}^{0i}),\label{mpunto}\\
{P}^{i\prime}&=&\frac{2c^{2}}{15G}{\sigma}_{R}^{jl\prime}{\sigma}%
_{I}^{kl\prime}\epsilon ^{ijk}+\frac{1}{6}\phi _{2I}^{0j}\phi _{2R}^{0k}\epsilon ^{ijk}. \label{ppunto}
\end{eqnarray}
Now, from eqs. (\ref{momdip}) and (\ref{Ddot}) we obtain
\begin{equation}\label{momento}
MR^{i\prime}=P^{i}+\frac{8}{5\sqrt{2}c}{\sigma}_{R}^{ij\prime}P^{j}-\frac{Q}{3c}\phi _{2R}^{0i}-\frac{1}{3\sqrt{2}c}(\phi _{1R}^{0k}\phi_{2I}^{0j}+\phi _{1I}^{0j}\phi_{2R}^{0k})\epsilon ^{ijk}.
\end{equation}
Finally, taking one more Bondi time derivative of (\ref{momento}) yields the equation of motion of the center of mass,
\begin{eqnarray}\label{ecmov}
M{R}^{i\prime\prime} &=&\frac{2c^{2}}{15G}{\sigma}_{R}^{jl\prime}{\sigma}%
_{I}^{kl\prime}\epsilon ^{ijk}+\frac{8}{5\sqrt{2}c}{\sigma}_{R}^{ij\prime\prime}P^{j}-\frac{Q}{3c}%
{\phi}_{2R}^{0i\prime} \nonumber\\
&&+\frac{1}{6}\phi _{2I}^{0j}\phi _{2R}^{0k}\epsilon ^{ijk}-\frac{1}{3\sqrt{2}c}%
(\phi _{1R}^{0k}\phi _{2I}^{0j}+\phi _{1I}^{0j}\phi _{2R}^{0k})^{\prime}%
\epsilon ^{ijk}.
\end{eqnarray}
The r.h.s. of the equation only depends on the gravitational data, the initial mass of the system, and the electromagnetic radiation at null infinity. Now, we assume that the main contribution in the Maxwell scalars is coming from the monopolar and the electromagnetic dipole moment.
In a Bondi frame, the electromagnetic dipole moment for an asymptotically flat spacetime is given by \cite{ANK}
\begin{equation}\label{cq1}
p^{i}+ic^{-1}\mu^{i} =\frac{1}{2}\phi _{0}^{0i},
\end{equation}
It is possible to extend this definition to a NU frame writing the following equation,
\begin{equation}\label{cq2}
p^{\ast i}+ic^{-1}\mu^{\ast i} =\frac{1}{2}\left[ \frac{\phi _{0}^{0\ast }}{Z^{\prime 2}}\right]^{i}.
\end{equation}
It is quite convenient to write the Maxwell field at infinity in terms of $p^i$ and $\mu^i$. For that, we need to solve the linear part of the scalars $\phi _{2}^{0i},\phi _{1}^{0i}$, and $\phi _{0}^{0i}$ because all approximations are up to the second order in Maxwell and gravitational data. Also, we need the quadratic contribution only for $Q{\phi}_{2R}^{0i\prime}$ of this equation. So, inserting the constant $G$, $c$, and the $\sqrt{2}$ factor of the retarded time in the linearized Bianchi identities (eqs. \ref{bianchi4} and \ref{bianchi3}), we can write
\begin{eqnarray*}
\phi _{1}^{0i} &=&\frac{{\phi}_{0}^{0i\prime}}{\sqrt{2}c}, \\
\phi _{2}^{0i} &=&-\frac{\sqrt{2}{\phi}_{1}^{0i\prime}}{c},
\end{eqnarray*}
Finally, using the definition for the center of charge we have
\begin{align}
\phi _{0R}^{0i} &=2p^i, \quad \phi _{0I}^{0i} =\frac{2\mu^{i}}{c},  \nonumber\\
\phi _{1R}^{0i} &=\frac{2{p}^{i\prime}}{\sqrt{2}c}, \quad \phi _{1I}^{0i} =\frac{2{\mu}^{i\prime}}{\sqrt{2}c^2}, \label{MaxLin} \\
\phi _{2R}^{0i} &=-\frac{2 p^{i\prime \prime}}{c^{2}}, \quad \phi _{2I}^{0i} =-\frac{2{\mu}^{i\prime\prime}}{c^{3}}. \nonumber
\end{align}
The quadratic contribution needed in the r.h.s of eqs. (\ref{Ddot}), (\ref{momento}), and (\ref{ecmov}) can be obtained expanding the Bianchi equation (\ref{bianchi3}) in the following way,
\begin{equation}
\phi_{0R}^{0i\prime\prime}=-c^2\phi _{2R}^{0i}+\frac{3c}{5\sqrt{2}}(\sigma_{R}^{ij}\phi _{2R}^{0j}-\sigma _{I}^{ij}\phi _{2I}^{0j})^{\prime},
\end{equation}
and using the linear Maxwell fields (\ref{MaxLin}) in the last equation, we get
\begin{equation}
\phi _{2R}^{0i}=-\frac{2p^{i\prime\prime}}{c^2}-\frac{3}{c5\sqrt{2}}(\sigma_{R}^{ij}{p}^{j\prime\prime}-c^{-1}\sigma _{I}^{ij}{\mu}^{j\prime\prime})^{\prime}. \label{dDD}
\end{equation}
Now, we are ready to write the equation of motion of the center of mass staring with  eq. (\ref{ecmov}), and combining this with eq. (\ref{MaxLin}-\ref{dDD}), we can therefore write
\begin{eqnarray}
M{R}^{i\prime \prime } &=&\frac{2c^{2}}{15G}{\sigma }_{R}^{jl\prime }{\sigma
}_{I}^{kl\prime }\epsilon ^{ijk}+\frac{8}{5\sqrt{2}c}{\sigma }_{R}^{ij\prime
\prime }P^{j} \nonumber\\
&&+\frac{2Q}{3c^{3}}p^{i\prime \prime \prime}+\frac{Q}{c^{2}5\sqrt{2}}(\sigma
_{R}^{ij}{p}^{j\prime \prime }-c^{-1}\sigma _{I}^{ij}{\mu }^{j\prime \prime
})^{\prime } \nonumber\\
&&+\frac{2}{3c^{5}}{\mu }^{j\prime \prime }p^{k\prime \prime }\epsilon
^{ijk}+\frac{2}{3c^{5}}({\mu }^{j\prime }{p}^{k\prime })^{\prime \prime
}\epsilon ^{ijk}.  \label{cofmE}
\end{eqnarray}
The electromagnetic fields, including in (\ref{cofmE}), generalise the equation of motion of the center of mass. Also, there are four kinds of terms, a gravitational radiation term, the classical electrodynamic radiation reaction force, an electro-gravity coupling, and a pure electromagnetic force.

\section{Comparison with ANK equations}
In a previous work \cite{KQ2}, the difference between Kozameh-Quiroga (KQ) and Adamo-Newman-Kozameh (ANK) equations has been discussed; so in this section we give a summary of these differences and we compare the equation of motion of the center of mass when electromagnetic contributions are present. Before that, we list the main differences between KQ and ANK formalisms
\begin{itemize}
  \item Our definition of angular and mass dipole momenta used the Geroch-Winicour-Tamburino linkages \cite{Wini,GW}, whereas the ANK uses the linear part of the linkage, i.e. $[\psi_1^0]^i$
  \item The cut equations, i.e. the equation relating the point in the space with cuts in $\scri$ in the ANK approach is the good cut equation (\ref{good_cut}). The solution space of the good cut equation is a complex manifold, while KQ uses non vanishing shears obtained from the RNC cut equation (see sec. \ref{sec:RNC}) the solution space of the RNC cut equation is real.
  \item ANK defines the intrinsic angular momentum as the imaginary part of a complex worldline while KQ evaluates the angular momentum coming from the imaginary part of the linkage on the center of mass to define the spin.
\end{itemize}
First we introduce the mass dipole moment, angular momentum and spin definitions given in the ANK formalism \cite{ANK}
\begin{eqnarray}
D_{\text{\tiny\emph{ANK}}}^{i}&=&-\frac{c^{2}}{6\sqrt{2}G}\psi _{1R}^{0i},\\
J_{\text{\tiny\emph{ANK}}}^{i}&=&-\frac{c^{3}}{6\sqrt{2}G}\psi _{1I}^{0i},\\
S_{\text{\tiny\emph{ANK}}}^{i}&=&cM\xi _{I}^{i}.
\end{eqnarray}
Now computing the component $l=1$ of eq. (\ref{DJB}) we can write
\begin{eqnarray}
D^{i}&=&-\frac{c^{2}}{6\sqrt{2}G}\psi _{1R}^{0i}+\frac{c^{2}}{5G}\sigma
_{R}^{jl}\sigma _{I}^{kl}\epsilon ^{ijk} +  \mbox{\small higher harmonics}\nonumber\\
J^{i}&=&-\frac{c^{3}}{6\sqrt{2}G}[\psi _{1}^{0}-\sigma^{0}\eth \bar{\sigma}^{0}-\frac{1}{2}\eth(\sigma ^{0}\bar{\sigma}^{0})]_{I}^{i}.
\end{eqnarray}
The mass dipole moments are different in both formalisms, however the Bondi 4-momentum $P^{\alpha}$, and the electrodynamic dipole moment are the same in both formulations.
The equation of motion of the center of mass in the ANK equations is given by
\begin{eqnarray*}
M\xi _{R}^{i\prime \prime } &=&\frac{2\sqrt{2}c^{2}}{15G}{\sigma }%
_{R}^{jl\prime }{\sigma }_{I}^{kl\prime }\epsilon ^{ijk}-\frac{c^{2}}{G}%
(\sigma _{R}^{jl}\sigma _{I}^{kl})^{\prime \prime }\epsilon ^{ijk}-\frac{4}{5%
\sqrt{2}c}{\sigma }_{R}^{ij\prime \prime }P^{j} \\
&&+\frac{2Q}{3c^{3}}p^{i\prime \prime }+\frac{Q}{c^{2}15\sqrt{2}}(7{\sigma }%
_{R}^{ij}p^{j\prime \prime }-c^{-1}{\mu }^{j\prime \prime }{\sigma }%
_{I}^{ij})^{\prime } \\
&&+\frac{4}{3c^{5}}{\mu }^{j\prime \prime }p^{k\prime \prime }\epsilon
^{ijk}+\frac{2}{3c^{5}}\left( {\mu }^{j\prime }p^{k\prime }\right) ^{\prime
\prime }\epsilon ^{ijk} \\
&&-\frac{Q}{3c^{4}}\left( 3p^{k\prime \prime }\frac{S_{\text{\tiny\emph{ANK}}}^{k}}{cM}-{\mu }%
^{j\prime \prime }\xi _{R}^{k\prime \prime }\right) ^{\prime \prime
}\epsilon ^{ijk}
\end{eqnarray*}
where $\xi _{R}$ is the position of the center of mass, and $\xi _{I}$ is the imaginary part of the complex worldline. Now directly from eq. (\ref{cofmE}) we have
\begin{eqnarray*}
M{R}^{i\prime \prime } &=&\frac{2c^{2}}{15G}{\sigma }_{R}^{jl\prime }{\sigma
}_{I}^{kl\prime }\epsilon ^{ijk}+\frac{8}{5\sqrt{2}c}{\sigma }_{R}^{ij\prime
\prime }P^{j} \\
&&+\frac{2Q}{3c^{3}}p^{i\prime \prime }+\frac{Q}{c^{2}5\sqrt{2}}(\sigma
_{R}^{ij}{p}^{j\prime \prime }-c^{-1}\sigma _{I}^{ij}{\mu }^{j\prime \prime
})^{\prime } \\
&&+\frac{2}{3c^{5}}{\mu }^{j\prime \prime }p^{k\prime \prime }\epsilon
^{ijk}+\frac{2}{3c^{5}}({\mu }^{j\prime }{p}^{k\prime })^{\prime \prime
}\epsilon ^{ijk}.
\end{eqnarray*}
In the same way, the evolution of spin in both approaches is different, in the ANK equations is given by
\begin{eqnarray*}
S_{\text{\tiny\emph{ANK}}}^{i\prime } &=&\frac{6c^3}{5G}(\sigma _{R}^{kl}{\sigma }_{R}^{jl\prime
}+\sigma _{I}^{kl}{\sigma }_{I}^{jl\prime })\epsilon ^{ijk}+\frac{2}{3c^{3}}(%
{p}^{j\prime \prime }p^{k\prime }+c^{-1}{\mu }^{j\prime }{\mu }^{k\prime
\prime })\epsilon ^{ijk}-\frac{2Q{\mu }^{i\prime \prime }}{3c^{3}} \\
&&-\frac{4}{5\sqrt{2}c}{\sigma }_{I}^{ij\prime }P^{k}-\frac{Q}{3c^{3}}(\xi
_{R}^{k}p^{j\prime \prime }+\frac{S_{\text{\tiny\emph{ANK}}}^{k}}{cM}{\mu }^{j\prime \prime
})\epsilon ^{ijk},
\end{eqnarray*}
while in our approach this equation can be written as
\begin{equation*}
S^{i\prime }=\frac{c^{3}}{5G}(\sigma _{R}^{kl}{\sigma }_{R}^{jl\prime
}+\sigma _{I}^{kl}{\sigma }_{I}^{jl\prime })\epsilon ^{ijk}+\frac{2}{3c^{3}}%
(p^{j\prime \prime }{p}^{k\prime }+c^{-1}{\mu }^{j\prime }{\mu }^{k\prime
\prime })\epsilon ^{ijk}-\frac{2Q{\mu }^{i\prime \prime }}{3c^{3}}
\end{equation*}
Thus, it is interesting to see that the final equations in these two formulations have some similarities and many differences. These differences are a consequence of the definitions used in both formulations and the cuts equations. Additionally, these differences can be traced back in the ANK formulation to the use of the relation $\Psi^{ij}=-\bar{\sigma}^{ij}$ in eq. (6.33) \cite{ANK}, this relationship contradicts eq. (\ref{psiprima}) since $\Psi^{ij\prime}=-{\bar{\sigma}}^{ij\prime}$ and ${\Psi}^{ij\prime}$ must be quadratic in ${\sigma}^{ij\prime}$. Thus, we can conclude that some equations in the ANK approach are incorrect.

\section{Applications} \label{sec:4}

\subsection{Center of Charge and Gravitational Charged Particle}\label{sec:2.3}
The concept of center of charge is introduced, assuming that a worldline described by the RNC cut equations (\ref{Z1apr}) exists, where the position function is the center of charge of the system. In this worldline, the electric dipole moment is equal to zero, and the function $R^i(u)$ will be labeled as $X^i(u)$ in order to avoid confusion with the center of mass position. As for the center of mass, we need to use the transformation law between the Maxwell scalars $\phi_0^{0\ast}$ constructed using the NU tetrad, and $\phi_0^0$ given by eq. (\ref{maxwell}). Now, making a Taylor expansion of the above equation, and expanding around $u$ we get
\begin{align}
\frac{\phi _{0}^{0\ast }}{Z^{\prime 2}} &=\phi _{0}^{0}(u+\delta u)-2L\phi _{1}^{0}(u+\delta u), \nonumber\\
&=\phi _{0}^{0}(u)+{\phi}_{0}^{0\prime}(u)\delta u-2\eth \delta u\phi _{1}^{0}(u),\nonumber\\
&=\phi _{0}^{0}(u)-\eth \phi _{1}^{0}(u)\delta u-2\eth \delta u\phi _{1}^{0}(u),
\end{align}
where in the last equality, we have used the Bianchi identity (\ref{bianchi3}). Now, we take the $l=1$ component of this equation using the harmonic expansion (\ref{exp}),
\begin{equation}\label{phi0i}
\left[ \frac{\phi _{0}^{0\ast }}{Z^{\prime 2}}\right] ^{i}=\phi
_{0}^{0i}-2QX^{i}+\sigma _{R}^{ij}\phi _{1}^{0j}-i\frac{\sqrt{2}}{2}%
X^{j}\phi _{1}^{0k}\epsilon ^{ijk}.
\end{equation}
In the center of charge worldline, the electric dipole momentum must vanish at $u=const$, i.e. $p^{\ast i} = 0$. In this special worldline, the magnetic dipole moment $\mu^{\ast i}$ will be called intrinsic magnetic moment $\mu^i_S$ and $\mu^{i}$ will be the total magnetic moment. So, splitting into the real and the imaginary parts and using eqs. (\ref{cq1}-\ref{cq2}) we can write the following,
\begin{align}
p^{i}&=QX^{i}-\frac{1}{2}\sigma _{R}^{ij}\phi _{1R}^{0j}-\frac{1}{2\sqrt{2%
}}X^{j}\phi _{1I}^{0k}\epsilon ^{ijk}, \label{dipolarelectrico}\\
\mu ^{i}&=\mu _{S}^{i}-\frac{1}{2}\sigma _{R}^{ij}\phi _{1I}^{0j}+\frac{1%
}{2\sqrt{2}}X^{j}\phi _{1R}^{0k}\epsilon ^{ijk}. \label{mu}
\end{align}
Now, we are ready to write the equation of motion for the gravitational charged particle (GCP). The basic idea to model the GCP, is to assume that the worldline of the center of mass and the worldline of the center of charge coincide, i.e $R^i=X^i$. Now, staring from  eq. (\ref{ecmov}), assuming $R^i=X^i$, and using the linear part of (\ref{dipolarelectrico}-\ref{mu}) we can write,
\begin{eqnarray*}
M{R}^{i\prime \prime } &=&\frac{2c^{2}}{15G}{\sigma }_{R}^{jl\prime }{\sigma
}_{I}^{kl\prime }\epsilon ^{ijk}+\frac{8}{5\sqrt{2}c}{\sigma }_{R}^{ij\prime
\prime }P^{j} \\
&&+\frac{2Q^{2}}{3c^{3}}R^{i\prime \prime \prime }+\frac{Q}{c^{2}5\sqrt{2}}%
(\sigma _{R}^{ij}Q{R}^{j\prime \prime }-c^{-1}\sigma _{I}^{ij}{\mu }_S
^{j\prime \prime })^{\prime } \\
&&+\frac{2Q}{3c^{5}}{\mu }^{j\prime \prime }_S R^{k\prime \prime }\epsilon
^{ijk}+\frac{2Q}{3c^{5}}({\mu }^{j\prime }_S{R}^{k\prime })^{\prime \prime
}\epsilon ^{ijk}. \label{ALfinal}
\end{eqnarray*}
In this case, the first term of the r.h.s is exactly the contribution to the momentum that yields the classical Abraham–Lorentz reaction force of classical electrodynamics \cite{landau1971classical}. It is well known that the Abraham-Lorentz recoil force have pathological solutions in which a particle accelerates in advance of the application of a force, these are the so-called pre-acceleration or runaway solutions. However, the equation (\ref{ALfinal}) adds some extra terms that depend on the Bondi free data, which would give well behaved solutions to the differential equations, i.e. solutions in which the acceleration goes to zero as time go to infinity.

\subsection{Gyromagnetic ratio of a charged black hole}
The study of the binary coalescence and their collapse into a black hole has regained an important relevance in General Relativity and astrophysics given the recent detection of gravitational waves made by LIGO \cite{LIGO2016}. In this context, as a second application of our approach, we propose to find the gyromagnetic ratio for Kerr-Newman black hole. In this analysis, we have to use the angular momentum (\ref{angmomentum}) and the magnetic moment (\ref{mu}), where the ratio magnetic moment-spin for a charged stationary black hole can be written as
\begin{equation}
\frac{\mu _{S}^{i}}{S^i}=\frac{\mu ^{i}}{J^i}
\end{equation}
We compute the $l=1$ component of $\phi_0^0$ and $\psi_1^0$ using the scalars of ref. \cite{KNblackhole}.  Then, the gyromagnetic ratio for stationary charged black hole is given by
\begin{equation}
\mu _{S}^{i}=g \frac{Q}{2M}S^i.
\end{equation}
where $g=2$ is the classical Landé g-factor, thus we have obtained the Dirac value of the gyromagnetic ratio \cite{newman2002classical} using the asymptotic fields at null infinity.

\section{Final remarks}

We have used the notion of center of mass and spin for asymptotically flat spacetime to derive a set of equations for gravitational isolated systems which emits gravitational radiation. Also, using the Maxwell fields received at infinity, we have found the electrodynamic contribution to the center of mass and spin for any asymptotically flat Einstein–Maxwell spacetime. The main ingredients used in this construction are the Newman-Penrose formalism, and the linkage integral together with a canonical NU foliation constructed from solutions to the regularized null cone cut.

We have compared our approach with the ANK formulation to check for differences and similarities. This comparison suggests that our equations, and definitions, are better suited to describe very energetic processes where an amount of energy is emitted as gravitational waves. Finally, to conclude this work, we used the notion of center of charge to build a simple model for a gravitational charged particle. In this construction, it is assumed that the center of charge worldline and the center of mass coincide. Additionally, it has been shown that for a Kerr-Newman black hole, the gyromagnetic ratio, i.e. the ratio of the magnetic moment-spin have the classical Landé value of $g=2$.

Finally, it can be mentioned that, the equations of motion obtained here could be used in many astrophysical situations to predict the motion of the center of mass from the emitted radiation or to predict the amount of radiation if the velocity and acceleration of the center of mass is given. This could be the case in closed binary coalescence, head-on collisions or supernova explosions, as long as, the velocity of the center of mass is low, compared to the speed of light. Additionally, the formalism could be generalized assuming relativistic velocities and releasing some of the work assumptions. However, several of these ideas we will be exploring in future works.

\begin{acknowledgements}
The work presented has been supported by VIE-UIS and the postdoctoral research program RC N001-1518-2016.
\end{acknowledgements}

\bibliographystyle{spmpsci}      



%

\end{document}